\documentclass[12pt]{book} 
\usepackage{myplenum}
\newcommand{\cyte}[1]{\refnote{\cite{#1}}}
\newcommand{\QKD}{{\sc qkd}}
\newcommand{\PDC}{{\sc pdc}}
\newcommand{\WCP}{{\sc wcp}}
\newcommand{\cancel}[1]{}
\begin{document}

\chapter{QUANTUM CRYPTOGRAPHY\\VIA PARAMETRIC DOWNCONVERSION}
\author{Gilles Brassard,\refnote{1} Tal Mor,\refnote{1,2}
and Barry C.\ Sanders,\refnote{3}}

\affiliation{\affnote{1}D\'epartement IRO, Universit\'e de Montr\'eal\\
C.P. 6128, succ.~centre--ville, Montr\'eal (Qu\'ebec), Canada H3C 3J7\\
email: {\tt \{brassard,mor\}@iro.umontreal.ca}\\[1mm]
\affnote{2}Electrical Engineering, University of California at Los Angeles\\
Los Angeles, CA 90095--1594, USA\\
email: {\tt talmo@EE.UCLA.EDU}\\[1mm]
\affnote{3}Department of Physics, Macquarie University\\
Sydney, New South Wales, Australia 2109\\
email: {\tt barry@mpce.mq.edu.au}
}

\begin{abstract}

The use of quantum bits (qubits) in cryptography holds the promise 
of secure cryptographic quantum key distribution schemes.
It is based usually
on single-photon polarization states. 
Unfortunately, the implemented ``qubits'' in the usual weak pulse
experiments are not true two-level systems, and quantum key distribution based
on these imperfect qubits is totally insecure in the presence of high 
(realistic) loss
rate.  In~this work, we investigate another potential implementation:
qubits generated \mbox{using} a process of parametric downconversion.
We find that, to first (two-photon) and second (four-photon)
order in the parametric downconversion small parameter,
this implementation of quantum key distribution is equivalent to
the theoretical version.

Once realistic measurements
are taken into account, quantum key distribution based on parametric
downconversion suffers also from sensitivity to extremely high 
(nonrealistic) losses.
By choosing the small parameter of the process according to the loss rates,
both implementations of quantum key distribution can in principle
become secure against
the attack studied in this paper.  However, adjusting the small 
parameter to the required levels 
seems to be impractical in the weak pulse process.
On the other hand, this can easily be done in the parametric 
downconversion process, making it a much more promising implementation.

\end{abstract}

\vspace{5mm}
\noindent
pacs: 03.67.Dd, 42.50.Ar, 42.50.Dv, 03.65.Bz, 89.80.+h

\section{INTRODUCTION}

Quantum information theory suggests the possibility to accomplish tasks
that are beyond the capability of classical computer science,
such as 
information-secure cryptographic key distribution\cyte{BB84}. 
While theoretical quantum key distribution (\QKD{}) 
schemes are proven secure against very sophisticated 
attacks\cyte{BBBGM}, the experimental \QKD{} schemes are not yet proven
secure even against very simple attacks.
In this work, we analyse the effect of losses on the security of 
experimental quantum key distribution.
We investigate a novel implementation,
qubits produced by a process of parametric downconversion (\PDC{}),
and we compare it to the more common implementation based on weak coherent
pulses (\WCP{}). 

A protocol is considered secure if 
the adversary is restricted only by the rules of quantum mechanics,
and yet cannot obtain any information on the final key.
In the four-state scheme\cyte{BB84} usually referred to as BB84,
the sender (Alice) and the receiver (Bob) use 
two conjugate bases (say, the rectilinear basis, $+$, and the diagonal
basis, $\times$) for the polarization of single photons.
In basis $+$ (resp.~$\times$), they use the two orthogonal basis states
$|0_+ \rangle$ and $|1_+ \rangle $ 
(resp.~$|0_\times \rangle$ and $|1_\times \rangle $)
to represent ``0'' and ``1'' respectively. 
The basis is revealed later on, which enables Bob to decode the bit whenever
he used the same basis as Alice; otherwise, they throw the bit away.
Finally, they use error-correction and privacy amplification to obtain
a potentially secure final key\cyte{BBBSS,BBBGM}.
\looseness=-1

All the experiments done so far to demonstrate protocols for secure
quantum key distribution use pulses of light containing (on average)
much less than one photon. 
We~approximate the state of the
modified qubit created by this process to be in
single mode, which we call 
a ``weak coherent pulse'' (\WCP{}).
[For an explanation regarding a description of a pulse, see Blow et 
al\cyte{pulses}.]
We analyse the security of \WCP{}-based schemes while
paying special attention to the losses.
The channel causes huge loss rate
(whether a fiber, which causes attenuation, or free space, which causes
beam broadening).
In~the experimental literature, 
it is usually assumed that the only effect of losses
is to reduce the bit rate. 
We show that there are two different types
of losses, channel losses and losses due to the state (``state losses'').
The state losses have impact on the bit rate.
The channel losses
have a vital impact on security, in
addition to their impact on the bit rate.
A~careful analysis of channel losses shows that schemes that
were assumed secure are in fact totally insecure even against a simple 
intercept-resend attack.
In~intercept-resend attacks, an eavesdropper (Eve)
performs a complete measurement on the input qubit, and she prepares and
sends to Bob a state of her own, according to the outcome of her measurement.
When Alice and Bob are using linearly independent states,
Eve can {\em sometimes} get full information 
by performing a ``positive operator value measure'' (POVM)
that conclusively distinguishes such states.
This is fatal in presence of high channel losses between Alice and Bob because
Eve can recreate the state near Bob and send it to him without loss whenever
she measured it conclusively, whereas she forwards nothing to Bob otherwise!
We shall refer to this attack as the {\em conclusive-measurement attack}.
This was discussed when the two-state scheme\cyte{Ben92} was invented,
and its power against the four-state scheme
was realized  by Yuen\cyte{Yuen}.

Recently, parametric downconversion has been used to generate a 
polarization singlet state\cyte{OM} to test Bell's inequalities, 
and it is believed that it can be used
as a much better single-photon source for quantum key distribution.
Here, we explain the potential experiment
and we present the modified singlet state resulting from this \PDC{} process.
Then, we calculate the state sent to Bob, including
two-photon and four-photon terms, assuming dispersion-free devices,
no dark counts and perfect detectors.
[A different use of a \PDC{} for \QKD{} was previously suggested\cyte{ERTP},
based on Franson-type uncertainties,
but the polarization encoding we suggest here allows for a much 
simpler analysis].

We find that \PDC{}-\QKD{} is much more secure than \WCP{}-\QKD{}:
The security of \WCP{}-\QKD{} is destroyed in the presence of high channel
loss rate due to the linear independence
obtained when adding the second-order terms.
The crucial advantage of the \PDC{}-\QKD{} is that the second-order
terms do not affect the fact that the states in one basis
are linearly dependent on the states in the other basis.
Thus, the attack that destroys the security of \WCP{}-\QKD{} in the presence
of high losses has no impact on \PDC{}-\QKD{} (when second-order
calculation and perfect detection are considered).

When imperfections in the process 
are taken into account, this euphoric picture changes,
and the second-order states sent to Bob
are not linearly dependent anymore.
Fortunately, \PDC{}-\QKD{} becomes totally insecure
against the conclusive-measurement attack only in the presence of 
such extremely high loss rate that more serious practical problems
would have already arisen, such as the importance of dark counts,
or errors due to various inaccuracies in the devices.
As we explain in the discussion, it is probably impossible to
make the \WCP{} implementation secure against the conclusive measurement 
attack, thus we suggest that the experimental effort should be directed
towards the implementation of \PDC{}-\QKD{}.

\section{SECURITY OF WCP-BASED QKD}

Experimental \QKD{} is mainly based on the use of weak pulses of coherent light.
By~definition, a pulse consists of a linear superposition of many frequency
contributions, but the laser pulse itself can be considered
to be in a single, localized mode provided that 
dispersion is
not significant in any of the optical elements\cyte{pulses}.

Using Fock state notation, $|0,0\rangle$ denotes the vacuum state, and
the state $|n_\updownarrow,m_\leftrightarrow\rangle$, which
describes $n$ photons with vertical
polarization and $m$ photons with horizontal polarization,
is denoted more simply by $|n,m\rangle$.
Ideally, the four BB84 states should be
$\mbox{$| \! \updownarrow \rangle$} = |0_+ \rangle = |1,0\rangle$ and 
$\mbox{$| \! \leftrightarrow \rangle$} = |1_+ \rangle = |0,1\rangle$
in the $+$ basis, and 
\mbox{$|0_\times \rangle = (1/\sqrt{2}) [ |1,0\rangle + |0,1\rangle]$}
and $|1_\times \rangle = (1/\sqrt{2}) [ |1,0\rangle - |0,1\rangle]$ 
in the $\times$ basis.

Consider now a weak coherent pulse with parameter $\alpha$,
meaning that a photon would be detected with probability $\alpha^2$
if the pulse were measured by a perfect detector.
If~this pulse is polarized in the $+$ basis, the two states are simply,
to second order in $\alpha$,
\begin{eqnarray*}
|0_+^{wcp} \rangle &\approx& \bigg(1 - \frac{\alpha^2}{2} \bigg)|0,0\rangle +  
\alpha |1,0 \rangle + \frac{\sqrt{2} \alpha^2}{2} |2,0 \rangle  \nonumber  \\
|1_+^{wcp} \rangle &\approx& \bigg(1 - \frac{\alpha^2}{2} \bigg)|0,0\rangle +  
\alpha |0,1 \rangle + \frac{\sqrt{2} \alpha^2}{2} |0,2 \rangle \ .
\end{eqnarray*}
However, the two states in the $\times$ basis, when expressed as Fock states
in terms of the $+$ basis, are more complicated:
\begin{eqnarray*}
|0_\times^{wcp} \rangle &\approx&
\bigg(1 - \frac{\alpha^2}{2} \bigg)|0,0\rangle +  
(\alpha/ \sqrt{2}) \Big[ |1,0 \rangle + |0,1 \rangle \Big]
+ \frac{\sqrt{2} \alpha^2}{4}
\Big[ |2,0 \rangle + \sqrt{2}  |1,1\rangle + |0,2 \rangle \Big]  \nonumber  \\
|1_\times^{wcp} \rangle &\approx&
\bigg(1 - \frac{\alpha^2}{2} \bigg)|0,0\rangle +  
(\alpha/ \sqrt{2}) \Big[ |1,0 \rangle - |0,1 \rangle \Big]
+ \frac{\sqrt{2} \alpha^2}{4}
\Big[ |2,0 \rangle - \sqrt{2}  |1,1\rangle + |0,2 \rangle \Big] \ .
\end{eqnarray*}
We call those four states the {\em modified qubits}.  Note that they
are not two-level systems anymore but six-level systems, or {\em qu-hexits}.

If we considered only the first order in $\alpha$, as is usually done,
the four states would behave very much like the ideal BB84 states leading
us to the wrong conclusion that the protocol is secure!
However, when the second order is considered, the two states in one basis
are no longer linear combinations of the two states in the other basis.
As~noted by Yuen\cyte{Yuen}, this linear independence in the six-dimensional
Hilbert space creates a fatal flaw for BB84 in the presence of high losses.
These states can be distinguished conclusively by an appropriate POVM\@.
Such measurement yields no information about the state most of the time,
but sometimes it identifies it unambiguously.  As explained in the
introduction, this allows for a successful conclusive-measurement 
attack provided
the loss rate expected by Alice and Bob is sufficiently high.
To provide numerical analysis, one must find the states that form the
POVM\@. This is a cumbersome calculation and we leave it for the final paper.
However, it is clear that the success probability is of order~$\alpha^2$
(relative to the one-photon counts).
Therefore, with Eve getting a conclusive result with relative probability
of order $\alpha^2$, and with $\alpha^2=0.1$ as in the current experiments,
it seems that a channel loss rate of 90\%--95\% is fatal.
With current channel loss rates, there is no escape from decreasing
$\alpha$ by more than one order of magnitude if reasonable security 
is to be achieved, and by more than two orders of magnitudes if we expect
to have secure key distribution to distances required for practical purposes.

\section{CREATING A MODIFIED SINGLET STATE IN THE PROCESS OF PDC}

In this section, we present the parametric downconversion process
and we give the output state to second order in the \PDC{} parameter.
The \PDC{} process provides a source of
photons for Bob and Alice with important advantages over the weak coherent
pulse discussed in the previous section.
A~classical pump field with vertical polarization
drives a \PDC{} crystal below threshold,
thereby producing photon pairs from a
two-mode vacuum state input field $|0,0,0,0\rangle$.
The two output fields from the parametric downconverter are
correlated in time of emission as well as polarization,
and conservation laws apply to
the sum of energies and momenta of the photons in the two fields.
The~quantum field input to the parametric downconverter is
assumed to be in the vacuum state.
We consider the field emitted by the \PDC{} process
and channeled through a polarization rotator and a beam splitter,
which creates entanglement between them.
One arm of the resulting output goes to Alice and the other arm goes to Bob.

We denote by $|k_{a_\updownarrow},l_{a_\leftrightarrow},
n_{b_\updownarrow},m_{b_\leftrightarrow}\rangle$,
or more simply $|k,l,n,m\rangle$,
the state in which there are $k$ photons with vertical
polarization and $l$ photons with horizontal polarization
going into Alice's arm ``a'', 
and $n$ photons with vertical polarization
and $m$ photons with horizontal polarization
going into Bob's arm ``b''.
The \PDC{} small parameter~$\chi$, which is proportional to the strength 
of the pump field, the interaction time between the field and the crystal 
and the nonlinearity of the medium, is so that a photon pair 
would be detected with probability $\chi^2$ if the output of the
interaction were measured by perfect detectors.
The state created by this process is an entangled state,
and it is usually assumed to be a singlet $|\psi_-\rangle
= (1/\sqrt{2}) [| 0,1,1,0 \rangle - | 1,0,0,1 \rangle] $,
but we show in the final paper how to
calculate it more precisely, to obtain the {\em modified singlet\/}  
$ |\chi \rangle = |\psi_-^{(mod)} \rangle$ to second order in~$\chi$:
\begin{eqnarray}
\label{mod-singlet}
| \chi \rangle 
&=& 
\bigg(1 - \frac{\chi^2}{2} \bigg) |0,0,0,0\rangle
+ \frac{\chi}{2}\Big[
   |0,1,1,0\rangle
 + |1,1,0,0\rangle
 - |0,0,1,1\rangle
 - |1,0,0,1\rangle\Big]
\nonumber \\*
&&\mbox{} +  
\frac{\chi^2}{4}\Big[
   |0,2,2,0\rangle
 + |2,2,0,0\rangle
 + |0,0,2,2\rangle
 + |2,0,0,2\rangle
 - 2 |1,1,1,1\rangle
\nonumber \\*
&& \phantom{\mbox{} + \frac{\chi^2}{4}\Big[}
+ \sqrt{2} |1,0,1,2\rangle
 - \sqrt{2} |0,1,2,1\rangle
 + \sqrt{2} |1,2,1,0\rangle
 - \sqrt{2} |2,1,0,1\rangle \Big] 
 \ .
 \end{eqnarray}

\section{CREATING A MODIFIED QUBIT IN A PDC PROCESS}

In order to use \PDC{} for performing the BB84 four-state scheme,
we need to consider the state sent from Alice to Bob.
This is produced by Alice measuring
her arm in a basis ($+$ or $\times$) of her choice and letting the
other arm, which is the modified qubit, go to Bob through the
quantum channel.
More precisely, Alice directs her arm to an adjustable rotator (to choose the
basis of measurement: angle 0 for $+$ and $\pi/4$ for~$\times$) followed by a
polarization-dependent beam splitter that sends the horizontal mode to one
direction and the vertical mode to another spatial direction.
Each of these spatial modes is now subjected to a measurement,
which in the limit of perfect efficiency provides an exact count
of the number of photons that reached each detector.
In this section,
we analyse to orders $\chi$ and $\chi^2$ the modified qubit thus sent
to Bob resulting from the modified singlet state.

Considering $|\chi\rangle $ to order $\chi$ 
and perfect detectors (used by Alice), the modified singlet
is projected to yield
a perfect qubit that is sent towards Bob in one of the BB84 states.
With imperfect detection, but not allowing dark counts,
Alice might send the vacuum, while she
thinks she sent a single photon, but this causes only state losses
and it has no effect on security
as far as we could see.

When we consider $|\chi\rangle $ to order $\chi^2$ 
and perfect detectors,
this process yields a modification of the four BB84 states,
but surprisingly still results in a perfect BB84 scheme!
With perfect measurements, only the terms with exactly
one photon at Alice's site will not be discarded,
so that we need only consider the terms 
$\frac{\chi}{2}[
   |0,1,1,0\rangle
 - |1,0,0,1\rangle]$ and 
$ \frac{\chi^2}{2\sqrt{2}}[
 |1,0,1,2\rangle - |0,1,2,1\rangle ] $.
In case Alice decides to rotate her mode by angle $\pi/4$ in order to send
Bob a qubit in the $\times$ basis, the above terms 
change to 
$\frac{\chi}{2\sqrt{2}}[
   |0,1,1,0\rangle
 + |1,0,1,0\rangle
 - |1,0,0,1\rangle
 + |0,1,0,1\rangle]$ and 
$\frac{\chi^2}{4}[
   |1,0,1,2\rangle 
 - |0,1,1,2\rangle 
 - |0,1,2,1\rangle  
 - |1,0,2,1\rangle ] $.

With ideal detectors and Alice measuring without rotation, the
state of Eq.~\ref{mod-singlet} is projected onto 
$|0,1\rangle$ or $|1,0\rangle$ (in Alice's arm), yielding respectively 
\begin{eqnarray*}\label{pdc-mod-qubit+}
|0_+^{pdc} \rangle &\approx& 
 \frac{\chi}{2} |1,0\rangle - \frac{ \chi^2}{2\sqrt{2}} |2,1\rangle
\\
|1_+^{pdc} \rangle &\approx& 
 - \frac{\chi}{2} |0,1\rangle + \frac{ \chi^2}{2\sqrt{2}} |1,2\rangle
\end{eqnarray*}
(since Alice used the $+$ basis). 
When Alice uses the $\times$ basis, the rotated terms (calculated as before) 
provide the relevant contribution, yielding
\begin{eqnarray*}\label{pdc-mod-qubitx}
|0_\times^{pdc} \rangle &\approx& 
 \frac{\chi}{2\sqrt{2}} \Big[|1,0\rangle + |0,1\rangle  \Big]
- \frac{ \chi^2}{4} \Big[ |1,2\rangle + |2,1\rangle \Big]
\\
|1_\times^{pdc} \rangle &\approx&
  \frac{\chi}{2\sqrt{2}} \Big[|1,0\rangle -|0,1\rangle \Big] +
 \frac{ \chi^2}{4} \Big[ |1,2\rangle - |2,1\rangle \Big] 
\ .
\end{eqnarray*}
The modified qubit is not a two-level system but a four-level
system. 
Yet, all four states lie in a two-level system spanned by any two of
them. Furthermore, they satisfy the same conditions as the theoretical 
BB84 states; each one in the $\times$ basis is an equal superposition
of the states in the $+$ basis.
Thus, all theoretical security analyses apply to these states.

\section{DISCUSSION}

We have seen that \PDC{}-\QKD{} has a crucial advantage over 
\WCP{}-\QKD{} due to the fact that the four states 
created in the \PDC{} process are equivalent 
to the theoretical states.
However, the calculation so far assumed that Alice uses perfect
measuring devices. 
A calculation taking account of realistic measurements will contain also 
other corrections.
Then the states {\em will be\/} linearly
independent, so that Eve can find a POVM to distinguish between them
conclusively.

Nevertheless, let us show a vital advantage of the more realistic
\PDC{}-\QKD{} over \WCP{}-\QKD{}.
Even though both schemes are insecure in principle in the presence of high
channel losses, the use of \PDC{} as
a source of qubits is potentially much preferable:
For \PDC{} qubits, the controlled parameter $\chi$ is usually
smaller than $10^{-3}$, thus the probability of having more than one
photon is $10^{-6}$, conditional to having at least one photon,
and seems to be negligible when the channel losses
are 99\% or even much more. Furthermore, the small parameter
can be easily further decreased according
to the loss rate to potentially solve the problem, 
perhaps while increasing the pulse frequency to keep the same bit rate.
In~\WCP{},
the corresponding parameter $\alpha$ is usually around~$0.3$.
Unfortunately, this parameter cannot be adjusted so easily because
it plays a dual role. Decreasing it immediately increases the state losses,
which are $1-\alpha^2$.
Although these are state losses and 
not channel losses---hence we didn't see any effect of these
losses on security---they are crucial in this implementation:
with much smaller $\alpha$ it is impossible to achieve any reasonable
bit rate since the state loss rate is $1-\alpha^2$.
Increasing the number of pulses to overcome this problem
is not an appropriate solution 
since Alice needs to write down the polarization of the
states in all pulses, and change the 
polarization for each one.

Another important advantage of \PDC{}-\QKD{} is that it solves
a problem usually left unnoticed: Eve can attack \WCP{}-\QKD{}
by eavesdropping {\em into Alice's lab\/};
this can be done
by finding the setting of Alice's polarizers using
a strong pulse sent to, and reflected from the polarizers\cyte{lab}
in between Alice's pulses. 
We are not aware of any such attack that can be used against the
\PDC{}-\QKD{} implementation.

Our work is only an initial step.
Analysis of more realistic scenarios and of other attacks
might show that \PDC{}-\QKD{} is not as superior to \WCP{}-\QKD{}
as this preliminary study indicates.

\section{ACKNOWLEDGEMENTS}
We are very thankful to Amiram Ron for helpful questions and remarks.
T.~Mor is thankful to Eugene Polzik for providing the initial motivation
for this work, and to the AQIP'98 conference, organized by
{\sc Brics} in Denmark.
B.~Sanders is thankful to S.~Warburton for checking calculations
and for useful comments.


\begin{numbibliography}
\bibitem {lab} Such an attack was suggested by several people
independently, such as Adi Shamir, Charles H. Bennett and others.
\end{numbibliography}
\end{document}